\documentclass[showpacs,amsmath,amssymb,aps,prd,groupedaddress,superscriptaddress,nofootinbib]{revtex4-2} 
\usepackage[utf8]{inputenc}
\usepackage{graphicx}
\usepackage{tensor}
\usepackage{bm}
\usepackage{color}
\usepackage{array}
\usepackage{tabularx}

\newcommand{\rsconnection}[1]{\tensor{\mathring{\omega}}{#1}}

\newcommand{\sconnection}[2][]{\tensor{#1\omega}{#2}}
\newcommand{\potential}[1]{\tensor{\Sigma}{#1}}
\newcommand{\pd}[1]{\tensor{\partial}{#1}}
\newcommand{\nablab}{\bm{\nabla}\!}

\newcommand{\e}[2][]{\tensor{#1{e}}{#2}}

\newcommand{\teta}[2][]{\tensor{#1\vartheta}{#2}}

\newcommand{\nonholon}[2][]{\tensor{#1{\Omega}}{#2}}

\newcommand{\bracket}[1]{\left<#1\right>}
\newcommand{\torsion}[1]{\tensor{T}{#1}}

\newcommand{\energy}[1]{\tensor{t}{#1}}

\begin{document}


\title{Angular momentum in the teleparallel equivalent of general relativity}



\author{J. B. Formiga}
\email[]{jansen@fisica.ufpb.br}

\author{Costa, R. D}

\affiliation{Departamento de Física, Universidade Federal da Paraíba, Caixa Postal 5008, 58051-970 João Pessoa, Pb, Brazil}



\begin{abstract}
In teleparallelism one is able to tackle the gravitational energy and angular momentum problems in a way that distinguishes this theory from other theories of gravity, such as general relativity. However, unlike the $4$-momentum, the quantity that is usually identified with a type of angular momentum does not have a clear interpretation. This problem is discussed, in particular the  vanishing of the $3$-angular momentum in the time gauge, and some general properties are obtained.
\end{abstract}


\maketitle

\section{Introduction} 
The importance of concepts such as energy and momentum has led to a variety of approaches to the gravitational energy problem, and the teleparallel equivalent of general relativity (TEGR) has played a distinguished role in providing a possible solution. The related problem of defining an angular momentum has also attracted the interest of many researchers and the TEGR offers an interesting candidate that does not rely on asymptotic flatness. Here we study this angular momentum, generalizing the approach that is usually used in the literature, and discuss its vanishing result in the time gauge with the help of G{\"odel}'s universe.

As is well known, the notion of energy is used in many areas of physics. It is import to characterize stationary states and solve equations of motions. It is also used in the study of gravitational waves \cite{MicheleMaggiore1}, black hole thermodynamics \cite{PhysRevLett.107.241301,PhysRevD.106.126007,Ma_2014}, and is crucial to statistical mechanics \cite{CRovelli1993}. However, there is no satisfactory way to define the local energy density of the gravitational field. Many proposals for the gravitational energy in the context of General Relativity (GR) have been made \cite{TheBerlinYearsVolume6p98,10.2307/20488488,PhysRev.89.400,Weinberg:1972kfs,Landaufourthv2,OZKURT2017}, but they fail to give a coordinate independent description. The TEGR, however, differs significantly from those approaches because it uses the notion of a tetrad field to define the conserved laws, and these laws and the gravitational energy-momentum tensor are independent of the coordinate system. Although these quantities depend on the tetrad field, it is possible to show that this dependency appears through the Levi-Civita connection coefficients of the tetrad field \cite{doi:10.1142/S0219887824502116}, which also happens with the acceleration tensor; so, in some sense, the teleparallel quantities have the same nature as the acceleration tensor and represent meaningful properties of both the observer and spacetime. (For interesting results obtained in the context of the TEGR, see Refs.~\cite{Gonalves2021,doi:10.1142/S021773232150125X,PhysRevD.106.044021,PhysRevD.108.044043,doi:10.1142/S0217732325501056,Carneiro_2025}.)

Like the gravitational energy, which is defined only asymptotically (Bondi-Sachs mass or ADM mass \cite{BONDI1960,doi:10.1098/rspa.1962.0161,1ddf44c6-0534-33d9-9b08-8c2c7c05245e,Arnowitt2008}, for example), the angular momentum in GR also suffers from severe limitations and difficulties \cite{Penrose1982,PhysRev.89.400}.  In the context of the TEGR, Maluf et al. have put forward a candidate for the angular momentum  that has some interesting properties \cite{PhysRevD.65.124001}: it appears naturally in the Hamiltonian formulation of the TEGR and is associated to a density that is defined at a spacetime point. (The latter property allows us to perform local and quasi-local calculations.) Furthermore, TEGR uses tetrads as the fundamental field and, as pointed out by some authors \cite{doi:10.1142/S021827181242014X,RovelliandVidotto}, we need tetrads to couple gravity to fermions, which means that they describe the gravitational field better.

In the next section we give a brief overview of teleparallelism, generalize Maluf et al. angular momentum by taking into account the formalism with an ``arbitrary'' connection, and obtain the conservation law for the angular momentum in a form that resembles that of special relativity. Then, we discuss the nature of the angular momentum. The difficulties in interpreting the angular momentum is analyzed from the perspective of different sets of observers in G{\"o}del's universe (Sec.~\ref{08012025a}). Finally, we conclude with some remarks in Sec.~\ref{26032025a}.

\section{Teleparallel field equations}
A theory is said to be teleparallel if the transport of a vector from one point to another is path independent, which is equivalent to saying that the curvature tensor of the affine connection used to transport the vector vanishes \cite{Eisenhart1927}. Although special relativity is a teleparallel theory, one usually uses the term ``teleparallelism'' to refer to theories whose metric cannot be reduced to the Minkowski metric by means of a coordinate transformation (nontrivial Levi-Civita connection). Therefore, the so-called teleparallel theories are endowed with an additional affine connection that can be made to vanish ``everywhere.''

This connection gives a variety of approaches that include different geometric settings and different action principles. One can formulate a teleparallel geometry with either torsion or nonmetricity (or both). For example, the case of a metrical connection with torsion was used by Einstein and corresponds to the most common approach to teleparallelism \cite{SAUER2006399,MOLLER1961118,PhysRevD.14.2521,PhysRevD.19.3524,ANDP:ANDP201200272,doi:10.1063/1.530774,PhysRevD.64.084014,Oko_w_2013,Oko_w_2014,PhysRevD.94.104045,Cai_2016,Blixt2021,PhysRevD.104.124074,Mashhoonbook2017}. Nonetheless, the case of a connection with nonmetricity has attracted the attention of several researches recently \cite{NesterPositive1989,PhysRevD.98.044048,PhysRevD.97.124025,doi:10.1142/S0219887823502158,AguiarGomes2023}. There are also two different ways to approach teleparallelism in the same geometric setting because, due to teleparallelism, we can always set the affine connection coefficients equal to zero. In one case, these coefficients are assumed to be zero right from the start \cite{ANDP:ANDP201200272}, in the action of the theory, while in the other case they are assumed to be arbitrary (see, e.g., Secs. 5.9 and 3.3 of Refs.~\cite{HEHL19951} and \cite{blagojevic2002gravitation}, respectively). Here, we refer to the former as the ``pure-tetrad'' approach and to the latter as the metric-affine gravity (MAG) approach.

We start with the field equations of the MAG approach and  limit ourselves to geometries with metric compatible connections and a nontrivial torsion. Then, we specialize to the pure-tetrad case.

Let $\{e_a\}$ and  $\{\teta{^a}\}$ denote the frame and the coframe, respectively; they satisfy the relation $\bracket{\teta{^a},\e{_b}}\equiv \teta{^a}(\e{_b})=\delta^a_b$, and their components in the coordinate bases $\{\partial_\mu\}$ and  $\{dx^\mu\}$ are $\tensor{e}{_a^\mu}$ and $\tensor{e}{^a_\mu}$. Greek letters are used to identify spacetime indices; Latin letters are used to denote $SO(1,3)$ indices, except for Latin letters in the middle of the alphabet ($i,j,k,\ldots$), which stand for spatial coordinate indices. The spacetime signature is $(-,+,+,+)$ and the Minkowski metric is denoted by $\eta_{ab}$. We assume $\{e_a\}$ is an orthonormal basis, i.e., $\eta_{ab}=\e{_a}\cdot\e{_b}$.

There exists a tetrad field $\e{_a}$ such that the teleparallel connection $\nablab$ (called the Weitzenb\"{o}ck connection) satisfies $\nablab_\mu\e{_a}=0$. This tetrad will be called the \textit{teleparallel frame} (TF). All teleparallel expressions can be written in terms of the TF, but we specialize to it only when dealing with the pure-tetrad approach.

The torsion components can always be written in the tetrad basis as
\begin{align}
\torsion{^a_b_c}=2\sconnection{^a_{[bc]}}+\nonholon{^a_b_c}, \label{16032019i}
\end{align}
where $\sconnection{^a_b_c}$ are the coefficients of the Weitzenb\"{o}ck connection in $\{\e{_a}\}$, defined as $\sconnection{^a_b_c}\equiv \bracket{\teta{^a},\nablab_b e_c }$, and $\nonholon{^a_b_c}\equiv-\bracket{\teta{^a},[e_b,e_c]}$ is the object of anholonomity. From the torsion tensor, we can define the so-called {\it torsion scalar} $T\equiv\potential{^a^b^c}\torsion{_a_b_c}$, where $\potential{^a^b^c}$ is called {\it superpotential} and defined as

\begin{align}
\potential{^\lambda^\mu^\nu}\equiv\frac{1}{4}\left( \torsion{^\lambda^\mu^\nu}+2\torsion{^{[\mu|}^\lambda^{|\nu]}}\right)+g^{\lambda[\nu}T^{\mu]}. \label{10112017l}
\end{align}
$g_{\mu\nu}$ is the spacetime metric and $\torsion{^\mu}\equiv\torsion{^\lambda_\lambda^\mu}$.

The Lagrangian density of a teleparallel theory that yields the same field equation for $g_{\mu\nu}$ as General Relativity is [see, e.g., Eq.~(3.61) of Ref.~\cite{blagojevic2002gravitation}]
\begin{align}
{\cal L}=-keT-c\tensor{\lambda}{_a_b^\mu^\nu}\tensor{R}{^a^b_\mu_\nu}-{\cal L}_M,
\end{align}
where  ${\cal L}_M$ is the matter Lagrangian density,  $e=\det||\e{^a_\mu}||$ is the determinant of the tetrad field, and $\tensor{\lambda}{_a_b^\mu^\nu}$ are Lagrange multipliers\footnote{They have the same symmetries as $\tensor{R}{^a^b_\mu_\nu}$:  $\tensor{\lambda}{_a_b^\mu^\nu}=-\tensor{\lambda}{_b_a^\mu^\nu}$ and $\tensor{\lambda}{_a_b^\mu^\nu}=-\tensor{\lambda}{_a_b^\nu^\mu}$. This means that there are $36$ independent Lagrange multipliers.}. Variations with respect to $\e{^a_\mu}$, $\sconnection{^a_b_c}$, and $\tensor{\lambda}{_a_b^\mu^\nu}$ gives [Eqs.~(3.63a) - (3.63c) of Blagojevi\'{c} \cite{blagojevic2002gravitation}] 
\begin{align}
\nabla_\alpha(\omega)\left( e\potential{^a^\mu^\alpha}\right)=\frac{e}{4k}\tau^{\mu a}, \label{29032019k}
\end{align}
\begin{align}
4\nabla_\nu(\omega) \tensor{\lambda}{_a_b^\mu^\nu}-\frac{1}{c}\tensor{M}{_a_b^\mu}=\tensor{\sigma}{^\mu_a_b},
\label{22102023a}
\\
\tensor{R}{^a^b_\mu_\nu}=0,
\end{align}
where
\begin{align}
\tau^{\mu a}=\energy{^\mu^a}+{\cal T}^{\mu a},
\\
\energy{^\mu ^a}=k(4\potential{^b^c^\mu}\torsion{_b_c^a}-\e{^a^\mu}T), \label{29032019h}
\end{align}
\begin{align}
\tensor{M}{^a^b^\mu}=-4ke\left(\potential{^a^\mu^b}-\potential{^b^\mu^a}\right),
\label{05062024c}
\end{align}
\begin{align}
\tensor{\sigma}{^\mu_a_b}=-\frac{1}{c}\frac{\delta {\cal L}_M}{\delta\sconnection{^a_\mu^b}},
\label{28102023a}
\end{align}
$k=c^4/(16\pi G)$, ${\cal T}^{\mu a}$ is the matter energy-momentum tensor, and the $\nabla_\nu(\omega)$ stands for the components of the covariant derivative that ``acts'' only on the $SO(1,3)$ indices\footnote{For example,  in Eq.~\eqref{22102023a}, the expression $\nabla_\nu(\omega) \tensor{\lambda}{_a_b^\mu^\nu}$ represents the components of $\nablab_\nu\left(\tensor{\lambda}{_a_b^\mu^\nu}\teta{^a}\otimes\teta{^b}\right)$: $\nabla_\nu(\omega) \tensor{\lambda}{_a_b^\mu^\nu}=\partial_\nu\tensor{\lambda}{_a_b^\mu^\nu}-\sconnection{^c_\nu_a}\tensor{\lambda}{_c_b^\mu^\nu}-\sconnection{^c_\nu_b}\tensor{\lambda}{_a_c^\mu^\nu}$.}.

Let us now prove an identity that we believe is not known in the literature. Using the property $\torsion{^a^b^c}=-\torsion{^a^c^b}$ and Eq.~\eqref{10112017l}, one can show that 
\begin{align}
\potential{^a^\mu^b}-\potential{^b^\mu^a}=(1/2)(\torsion{^\mu^a^b}+\e{^b^\mu}\torsion{^a}-\e{^a^\mu}\torsion{^b}).
\label{05032025b}
\end{align}
In turn, using the definitions $\nonholon{^\mu^a^b}\equiv \e{_c^\mu}\eta^{ad}\eta^{be}\nonholon{^c_d_e}$ and $\nonholon{^a}\equiv\eta^{ad}\nonholon{^c_c_d}$, we obtain
\begin{align}
e(\nonholon{^\mu^a^b}+\e{^b^\mu}\nonholon{^a}-\e{^a^\mu}\nonholon{^b})= -\partial_\alpha (\tensor{J}{^a^b^\mu^\alpha}/2k),
\label{05032025d}
\\
\tensor{J}{^a^b^\mu^\alpha}=2ke(\e{^a^\alpha}\e{^b^\mu}-\e{^a^\mu}\e{^b^\alpha}).
\label{23122024d}
\end{align}
Using Eqs.~\eqref{16032019i} and \eqref{05032025d} in Eq.~\eqref{05032025b} multiplied by $e$, we find 
\begin{align}
e\left(\potential{_a^\mu_b}-\potential{_b^\mu_a}\right)=-(1/4k)(\partial_\alpha \tensor{J}{_a_b^\mu^\alpha}
-\sconnection{^c_\alpha_a}\tensor{J}{_c_b^\mu^\alpha}
\nonumber\\
-\sconnection{^c_\alpha_b}\tensor{J}{_a_c^\mu^\alpha})
\label{05032025c}
\end{align}
The term in parenthesis on the right-hand side of this equation is exactly $\nabla_\alpha(\omega) \tensor{J}{_a_b^\mu^\alpha}$. Therefore, the tensor density $\tensor{M}{^a^ b^\mu}$, given by Eq.~\eqref{05062024c}, can be written as
\begin{align}
\tensor{M}{^a^b^\mu}=\nabla_\alpha(\omega) \tensor{J}{^a^b^\mu^\alpha}.
\label{20122024a}
\end{align}
As far as we know, only the particular case $\tensor{M}{^a^b^0}$ of Eq.~\eqref{20122024a} written in the TF is already known \cite{RochaNeto2014}.

Now we prove that $\tensor{M}{^a^b^\mu}$ is conserved and that the matter energy-momentum tensor has to be symmetric\footnote{This is a trivial result since ${\cal T}_{ab}$ is proportional to the Einstein tensor in the TEGR. An analogous proof was given by Obukhov and Pereira \cite{PhysRevD.67.044016} [see Eq.~(5.4) there].} in the TEGR. Since the connection $\sconnection{^a_b_c}$ has no curvature, the derivative $\nabla_\mu(\omega)\nabla_\alpha(\omega) \tensor{J}{_a_b^\mu^\alpha}$ commute. Thus, using the fact that the last two indices of $\tensor{J}{^a^b^\mu^\alpha}$ are antisymmetric, we find the conservation law
\begin{align}
\nabla_\mu(\omega)\tensor{M}{_a_b^\mu}=\nabla_\mu(\omega)\nabla_\alpha(\omega) \tensor{J}{_a_b^\mu^\alpha}=0.
\label{23122024a}
\end{align} 
To conclude that ${\cal T}_{ab}$ must be symmetric, we use $\nabla_\mu(\omega)\tensor{M}{_a_b^\mu}=e\left({\cal T}_{ba}-{\cal T}_{ab}\right)$ (see, e.g., p.~70 of Ref.~\cite{blagojevic2002gravitation}).  

The same argument above can be used to conclude that $\nabla_\mu(\omega)\nabla_\nu(\omega) \tensor{\lambda}{_a_b^\mu^\nu}=0$. Therefore, from Eq.~\eqref{22102023a}, we find that $\tensor{\sigma}{^\mu_a_b}$ is also conserved, i.e., $\nabla_\mu(\omega)\tensor{\sigma}{^\mu_a_b}=0$. [Note that if the matter Lagrangian does not depend on $\sconnection{^a_\mu_b}$, we will have $\tensor{\sigma}{^\mu_a_b}=0$. In this case, Eq.~\eqref{22102023a} becomes $\tensor{M}{_a_b^\mu}=4c\nabla_\nu(\omega)\tensor{\lambda}{_a_b^\mu^\nu}$.]

For comparison, we present the relation between our notation and that of Ref.~\cite{blagojevic2002gravitation} in Table \ref{28102023c}.

\begin{table}[h]
\centering
	\begin{tabularx}{0.45\textwidth}{|>{\centering\arraybackslash}X|>{\centering\arraybackslash}X|>{\centering\arraybackslash}X|}
	\hline
	 &Blagojevi\'c & Ours\\
	 \hline
	Spacetime indices & $\mu,\nu,\lambda,...$ (middle)& $\alpha,\beta, \gamma,...$ (whole)\\
	\hline
	Spatial spacetime indices & $\alpha,\beta,\gamma,...$ (beginning) & $i,j,k,...$ (middle)\\
	\hline 
	 $SO(1,3)$ indices & $i,j,k,...$ (middle)& $a,b,c,...$ (beginning)\\
	\hline
	Spatial $SO(1,3)$ indices & $a,b,c$ (beginning) & $(i),(j),(k),...$\\
	\hline
	Frame & $\tensor{h}{_i^\mu}$& $\e{_a^\mu}$\\
	\hline
	 Coframe & $\tensor{b}{^i_\mu}$& $\e{^a_\mu}$\\
	\hline
	 Affine connection & $\sconnection{^i_j_\mu}$	  & $\sconnection{^a_\mu_b}$\\
	\hline
	Signature & $-2$ & $+2$\\
	\hline
	Superpotential &$\beta^{ijk}$ & $(k/c)\potential{^a^b^c}$\\
	\hline
	Matter Energy-momentum & $ \tau^{\mu\nu}$ & $(e/c){\cal T}^{\mu\nu}$\\
	\hline
	\end{tabularx}
	\caption{Comparison between our notation and that of Blagojevi\'c \cite{blagojevic2002gravitation}.}
	\label{28102023c}
\end{table}

\subsection{Pure-tetrad approach}
The pure-tetrad approach is basically the case in which one works only with the TF\footnote{This does not mean that this approach is limited to only one TF, and no other can be used. If the new TF is related to the old one by a local Lorentz transformation, rather than a global transformation, one can change the TF by changing the affine connection, i.e., by choosing a different Weitzenb\"{o}ck geometry. For further discussion we refer to Ref.~\cite{doi:10.1142/S0219887824502116}.}. In that case the connection coefficients $\sconnection{^a_b_c}$ vanish. Hence, the torsion components \eqref{16032019i} become
\begin{align}
\torsion{^a_\mu_\nu}=\nonholon{^a_\mu_\nu}=\pd{_\mu}\e{^a_\nu}-\pd{_\nu}\e{^a_\mu}, \label{04102019p}
\end{align}
and the Levi-Civita connection can be obtained from
\begin{align}
\rsconnection{^a_b_c}=\frac{1}{2}\left( \torsion{_b_c^a}+\torsion{_c_b^a}-\torsion{^a_b_c} \right). \label{10092020a}
\end{align}
In turn, the field equations \eqref{29032019k} and \eqref{22102023a} read
\begin{align}
\pd{_\alpha}\left( e\potential{^a^\mu^\alpha}\right)=\frac{e}{4k}\left(\energy{^\mu^a}+{\cal T}^{\mu a}\right), 
\label{23122024c}
\\
\partial_\nu\left(4\tensor{\lambda}{_a_b^\mu^\nu}-\frac{1}{c}\tensor{J}{_a_b^\mu^\nu}\right)=\tensor{\sigma}{^\mu_a_b},
\label{07012025a}
\end{align}
where Eq.~\eqref{20122024a}, which becomes
\begin{align}
\tensor{M}{^a^b^\mu}=\partial_\alpha \tensor{J}{^a^b^\mu^\alpha},
\label{07012025b}
\end{align}
has been used. The superpotential $\potential{^a^b^c}$ can be given in terms of the Levi-Civita connection \cite{doi:10.1142/S0219887824502116}:
\begin{align}
\potential{_a_b_c}=\frac{1}{2}\rsconnection{_c_a_b}+\rsconnection{^d_d_{[c}}\tensor{\eta}{_{b]a}}.
\label{14082024h}
\end{align}

Equation \eqref{23122024c} is the usual field equation of the pure-tetrad approach (see, e.g., Maluf \cite{ANDP:ANDP201200272}). Although  Eq.~\eqref{07012025a} does not appear in this approach, we have chosen to write it here for completeness. (Some differences between the pure-tetrad approach and MAG may appear when $\tensor{\sigma}{^\mu_a_b}$ does not vanish.)

The quantity $\energy{^\mu^a}$ is interpreted as the ``gravitational energy-momentum tensor,'' while
\begin{align}
P^a_g\equiv\int_V d^3x e\energy{^0^a},\quad P^a_M\equiv\int_V d^3x e{\cal T}^{0a}
\label{02012023b}
\end{align}
can be interpreted as the gravitational and the matter $4$-momenta, respectively. Furthermore, we interpret the sum
\begin{align}
P^a\equiv P^a_g+P^a_M
\label{02012023a}
\end{align}
as the spacetime $4$-momentum. The integrals above are over a region $V$, a hypersurface defined by $x^0=$ constant. 

Maluf and others interpret the quantity $\tensor{M}{^a^b}\equiv \tensor{M}{^a^b^0}=\partial_\nu\tensor{J}{^a^b^0^\nu}$ as  the ``gravitational angular momentum density,'' and the quantity
\begin{align}
L^{ab}\equiv -\int_V d^3xM^{ab}
\label{30122019c}
\end{align}
as the ``total'' angular momentum of the gravitational field contained within $V$.

If the region $V$ is free from singularities, then, by using Eq.~\eqref{23122024c}, one finds that
 \begin{align}
P^a=4k\oint_S dS_i e\Sigma^{a0i},
\label{29102023c}
\end{align}
and, from [see Eqs.~\eqref{20122024a} and \eqref{23122024d}]
\begin{align}
M^{ab}=\tensor{M}{^a^b^0}=2k\partial_i\left[e\left(\e{^a^i}\e{^b^0}-\e{^b^i}\e{^a^0}\right)\right],
\label{19052024a}
\end{align}
one finds\footnote{There is a minus sign missing in Eq.~(85) of Ref.~\cite{https://doi.org/10.1002/andp.202300241}.} \cite{https://doi.org/10.1002/andp.202300241}
\begin{align}
L^{ab}=-\oint_S dS_i 2k\left[e\left(\e{^a^i}\e{^b^0}-\e{^b^i}\e{^a^0}\right)\right].
\label{19052024b}
\end{align}
[Note that, since we are using $k=c^4/(16\pi G)$ rather than $c^3/(16\pi G)$, the dimension of $L^{ab}$ defined here corresponds to that of Ref.~\cite{https://doi.org/10.1002/andp.202300241} multiplied by $c$.]

The quantity $\tensor{M}{_a_b^\mu}$ does not appear naturally in the pure-tetrad approach and has not been studied in detail in this context. Let us now show that $\tensor{M}{_a_b^\mu}$ is important to generalize the angular momentum approach in the TEGR. Equation \eqref{22102023a} already shows the relevance of this tensor. Nevertheless, we can go further and obtain a ``integral form of the law of conservation of angular momentum'' analogous to that of special relativity [see, e.g., Eq.~(5.48) of Ref.~\cite{Gravitation}]. Equation \eqref{23122024a} in the TF becomes
\begin{align}
\partial_\mu\tensor{M}{_a_b^\mu}=0.
\label{29102023d}
\end{align}
Therefore, we have
\begin{align}
\oint_{\partial \Omega} \tensor{M}{_a_b^\mu}dS_\mu=0,
\end{align}
where $\partial \Omega$ is the boundary of a four-dimensional region $\Omega$. This equation justifies the definition given by Eq.~\eqref{30122019c}, which corresponds to the integral over a spacelike surface of constant time. (Historically, the original motivation for $L^{ab}$ was one of the constraint equations of the Hamiltonian formulation of the TEGR \cite{ANDP:ANDP201200272}.)

\subsection{Interpreting $L^{ab}$}
\subsubsection{Angular momentum of what?}
The interpretation of $L^{ab}$ is not as clear as that of $P^a$, $P^a_g$ and $P^a_M$. For example, it is clear in Eq.~\eqref{02012023a} that $P^a$ can be seen as the sum of the gravitational $4$-momentum with the matter one. However, since Eq.~\eqref{22102023a} does not fix the Lagrange multipliers, the interpretation of the angular momentum given by Eq.~\eqref{30122019c} as the spacetime angular momentum (the sum of gravitation and matter angular momenta) or something else is not so clear.

One possible solution to this problem may be given by the relation $\partial_\mu\tensor{M}{_a_b^\mu}=e\left({\cal T}_{ba}-{\cal T}_{ab}\right)$, which is valid in the TF; it suggests that $L^{ab}$ is related to the matter fields, because of the similarity with the special relativistic case, where $\tensor{\mathcal{J}}{^\alpha^\beta^\gamma}=(x^\alpha-x_0^\alpha)\mathcal{T}^{\beta\gamma}-(x^\beta-x_0^\beta)\mathcal{T}^{\alpha\gamma}$ leads to $\partial_\gamma\tensor{\mathcal{J}}{^\alpha^\beta^\gamma}=\mathcal{T^{\beta\alpha}}-\mathcal{T}^{\alpha\beta}$. On the other hand, the Poincar\'e algebra obtained by Maluf \cite{ANDP:ANDP201200272} is satisfied by $L^{ab}$ and $P^a$; as $P^a$ is the spacetime $4$-momentum, it would be natural to assume that $L^{ab}$ is also a ``spacetime'' kind of angular momentum. By assuming that both views are right,  we are led to believe that $L^{ab}$ is both the spacetime and matter angular momentum, and that the contribution of the gravitational field is not given by a simple sum.

In classical field theory we have the canonical and the symmetric energy-momentum tensors, where the former is not necessarily symmetric. Without gravity, they lead to the same total energy and momentum, and the only difference is in the different description of where the energy is located. Once gravity is present, the location of the energy of the matter fields becomes important because this energy is the source of the gravitational field, and the nonsymmetric tensor cannot represent the real location of the matter energy. However, that requirement does not apply to the gravitational energy-momentum tensor. It turns out that $\energy{^a^b}$ is more like a canonical energy-momentum tensor. [In the context of the MAG, it coincides with the canonical energy-momentum three-form given by Eq.~(5.4.6) of Hehl et at. \cite{HEHL19951}]. Therefore, the location of the gravitational energy does not have to be frame independent and the condition $\energy{^a^b}=\energy{^b^a}$ is not necessary. The former can be verified by the fact the location of the gravitational energy predicted by $\energy{^\mu^a}$ depends on the chosen TF, while the latter can be verified from Eq.~\eqref{23122024c}: taking the antisymmetric part of  \eqref{23122024c} and using $\partial_\mu\tensor{M}{_a_b^\mu}=0$, we obtain\footnote{Due to the antisymmetry of the last two indices of $\potential{^a^\alpha^\mu}$, one can exchange $\partial_\alpha\e{^b_\mu}$ in Eq.~\eqref{24122024a} for $(1/2)\torsion{^b_\alpha_\mu}$.}
\begin{align}
\energy{^a^b}-\energy{^b^a}=-4k\left[\left(\partial_\alpha \e{^b_\mu}\right)\potential{^a^\alpha^\mu}-\left(\partial_\alpha \e{^a_\mu}\right)\potential{^b^\alpha^\mu}\right].
\label{24122024a}
\end{align}
It is not clear how the right-hand side of this equation can be associated to any sort of ``intrinsic'' angular momentum carried by the gravitational field, and how it affects $L^{ab}$.

Whatever the interpretation of $L^{ab}$, it seems clear that it is not an ordinary orbital angular momentum: it is hardy to believe that $\tensor{M}{_a_b^\mu}$ can reproduce  $\tensor{\mathcal{J}}{^\alpha^\beta^\gamma}=(x^\alpha-x_0^\alpha)\mathcal{T}^{\beta\gamma}-(x^\beta-x_0^\beta)\mathcal{T}^{\alpha\gamma}$ in some limit or approximation.

\subsubsection{Time gauge}
It has been shown that $L^{(i){(j)}}$, which yields the angular momentum \textit{per se}, vanishes in the time gauge \cite{PhysRevD.106.044021}. [This can be seen from Eq.~\eqref{19052024a}: the time gauge is characterized by $\e{_{(i)}^0}=0$, which leads to $M^{(i)(j)}=0$.] This raises some issues regarding the relation of $L^{(i){(j)}}$ with the TF. For example, why does $L^{(i){(j)}}$ vanish in $pp$-wave spacetimes regardless of the polarization either of the gravitational wave or the electromagnetic wave (see, e.g., Ref.~\cite{doi:10.1142/S0217732325501056})?  The frame where this happens is a freely falling frame that does not rotate and satisfies the time gauge, which is a type of TF that has yielded consistent results for $P^a$ \cite{doi:10.1142/S021773232150125X,Gonalves2021,PhysRevD.108.044043}.

Other examples are the Kerr and G{\"odel} spacetimes: the Gaussian coordinate systems used in Refs.~\cite{Novello2011} and \cite{Novello1993} allow us to find tetrad fields that are freely falling (but not necessarily without rotations) that satisfy the time gauge in Kerr and G{\"odel} spacetimes, respectively. It is hard to find a reasonable explanation for the vanishing of $L^{(i){(j)}}$ in those spacetimes. Not even the discussion in Ref.~\cite{doi:10.1142/S0217732322502224}, concerning the role of the constituents of the reference frame and their possible nongravitational interactions, can give us a satisfactory explanation. The reason for the latter point is this: some spacetimes may admit a time gauge for both a frame adapted to freely falling particles and a system of particles that are not free; they may also allow two distinct tetrads, both adapted to a system of freely falling particles, with one satisfying the time gauge and the other not. An example of the latter is the G{\"o}del spacetime: as mentioned before, any tetrad that is adapted to the coordinate system in Ref.~\cite{Novello1993} is adapted to a system of freely falling particles and satisfies the time gauge, hence  $L^{(i){(j)}}=0$; however, in Sec.~\ref{08012025a} we use a tetrad field that is adapted to free particles, but does not satisfy the time gauge and yield a nonvanishing $L^{(i){(j)}}$.

So, the nature of the system, whether or not is a system of free particles, and how they interact, is not enough to explain the behavior of $L^{abb}$. It is worth noting that this is not enough to explain the behavior of $P^a$ either. As discussed in Sec.~3.3.3 of Ref.~\cite{Gonalves2021}, the hypersurface of simultaneity also plays an important role. 

\section{G\"odel type spacetimes}\label{08012025a} 
The G\"odel universe is a cosmological solution of Einstein's field equations with rotating matter and a cosmological constant \cite{RevModPhys.21.447}. Although this spacetime is not supposed to represent any real cosmological model, it is frequently used for discussing philosophical issues and for studying the properties of solutions of GR \cite{Pfarr1981}. 

In order to discuss the angular momentum problem, we analyze two types of freely falling frames and an accelerated one. The first one does not satisfy the time gauge and gives predictions that are consistent with the rotation of the matter field; however, its predictions for $P^a$ are inconsistent. The second one will be adapted to the time gauge and, therefore, will have a vanishing angular momentum. The third is an accelerated frame that satisfies the time gauge and does not yield the same inconsistency as that of the first case.

\subsection{Freely falling without the time gauge}\label{19012025a}
Here we use a more general solution that includes G\"odel's one as a particular case. The line element of this  G\"odel-type solution can be written as\footnote{This metric can be obtained from the metric in Ref.~\cite{Fonseca-Neto1998} by changing the coordinates $t$ and $y$ there to $t-\sqrt{2}y$ and $\sqrt{2}y$, respectively; the signature is also different.} 
\begin{align}
ds^2=-\{cdt+\sqrt{2}[H(x)-1]dy\}^2+dx^2+H^2(x)dy^2+dz^2,
\label{09012025a}
\\
H(x)=e^{mx},\qquad m=\textrm{constant}.
\label{09012025b}
\end{align}
The constant $m$ is related to the cosmological constant $\Lambda$ and the rotation $\omega$ by $m^2=2\omega^2=-2\Lambda$. (Notice that this is a model with negative $\Lambda$.)

A frame that is freely falling but does not satisfy the time gauge is
\begin{align}
\e{_a^\mu}=-\hat{t}_a\hat{t}^\mu+\hat{s}_a\hat{s}^\mu+\hat{\phi}_a\hat{\phi}^\mu+\hat{z}_a\hat{z}^\mu,
\label{04032025b}\\
\e{^a_\mu}=-\hat{t}^a\hat{t}_\mu+\hat{s}^a\hat{s}_\mu+\hat{\phi}^a\hat{\phi}_\mu+\hat{z}^a\hat{z}_\mu,
\label{02062024c}
\end{align}
with
\begin{align}
\hat{t}^a=\delta^a_{(0)},\ \hat{t}_a=-\hat{t}^a,\ \hat{z}^a=\delta^a_{(3)},\ \hat{z}_a=\hat{z}^a,
\nonumber\\
\hat{s}^a\equiv \cos\phi\delta^a_{(1)}+\sin\phi\delta^a_{(2)},\ \hat{s}_a=\hat{s}^a, 
\nonumber\\
\hat{\phi}^a\equiv-\sin\phi\delta^a_{(1)}+\cos\phi\delta^a_{(2)},\ \hat{\phi}_a=\hat{\phi}^a,
\nonumber\\
\phi=\pi/2-mct/\sqrt{2},
\label{02062024d}
\end{align}
and
\begin{align}
\hat{t}^\mu=\delta^\mu_0,\ \hat{t}_\mu=-\delta_\mu^0-\sqrt{2}[H-1]\delta_\mu^2,
\nonumber\\
\hat{s}^\mu=\sqrt{2}[H^{-1}-1]\delta^\mu_0+H^{-1}\delta^\mu_2,\ \hat{s}_\mu=H\delta_\mu^2,
\nonumber\\
\hat{\phi}^\mu=-\delta^\mu_1,\ \hat{\phi}_\mu=-\delta_\mu^1,
\nonumber\\
\hat{z}^\mu=\delta^\mu_3,\ \hat{z}_\mu=\delta^3_\mu,
\label{02062024e}
\end{align}
where $\hat{t}^\mu\equiv\e{_a^\mu}\hat{t}^a$, $\hat{t}_\mu\equiv\e{^a_\mu}\hat{t}_a$ etc. (Note that $\hat{t}_a\equiv \eta_{ab}\hat{t}^b$, $\hat{s}_a\equiv \eta_{ab}\hat{s}^b$ etc.) The vectors $\{\hat{t},\hat{s},\hat{\phi},\hat{z}\}$ form an orthonormal basis in which $\hat{t}^a\hat{t}_a=-1$, $\hat{t}^a\hat{s}_a=0$, $\hat{s}^a\hat{s}_a=1$ etc.

The method used here to calculate the teleparallel quantities is basically that of Ref.~\cite{Formiga2021Braz} with a different metric signature. However, we will not use Eqs.~(28)-(31) there because they are written in a spherical coordinate system and, here, it is more convenient to perform the calculations in the coordinates $x^\mu=(ct,x,y,z)$.

Substituting Eq.~\eqref{02062024c} into Eq.~\eqref{04102019p} and using Eq.~\eqref{02062024e}, we obtain
\begin{align}
\torsion{^a_\mu_\nu}=2\sqrt{2}m\hat{t}^a\hat{s}_{[\mu}\hat{\phi}_{\nu]}-2m\hat{s}^a\left(\frac{\hat{t}_{[\mu}\hat{\phi}_{\nu]}}{\sqrt{2}}-\frac{\hat{s}_{[\mu}\hat{\phi}_{\nu]}}{H}\right)
\nonumber\\
+\frac{2m}{\sqrt{2}}\hat{\phi}^a\hat{t}_{[\mu}\hat{s}_{\nu]},
\label{02062024a}
\end{align}
where we have also used Eqs.~\eqref{03032025a} and \eqref{03032025b}. To obtain the Levi-Civita connection coefficients, we use Eq.~\eqref{02062024a} in Eq.~\eqref{10092020a}. This gives
\begin{align}
\rsconnection{_a_b_c}=m[-\hat{s}_b(\sqrt{2}\hat{t}_{[a}\hat{\phi}_{c]}+\frac{2}{H}\hat{s}_{[a}\hat{\phi}_{c]})+\sqrt{2}\hat{\phi}_b\hat{t}_{[a}\hat{s}_{c]}].
\label{02062024b}
\end{align}
From these coefficients, we can calculate all the relevant teleparallel quantities.

\subsubsection{Frame properties}
Let us analyze the properties of the observer congruence and frame. We first calculate the acceleration tensor and then decompose $\nablab_\mu\e{_{(0)}}$ into acceleration, \textit{vorticity}, and \textit{expansion tensor}.

Since the acceleration tensor can be given by $\tensor{\phi}{_a^b}=c\rsconnection{^b_{(0)}_a}=c\rsconnection{^b_c_a}\hat{t}^c$, we can easily see from Eq.~\eqref{02062024b} that the frame is freely falling with no rotation, i.e. $\tensor{\phi}{_a^b}=0$. This means that there is no nongravitational interaction in the frame.

The vector field $\e{_{(0)}}$ defines a congruence of curves, which corresponds to the worldline of the observers we are dealing with. A common way of studying a congruence with a tangent vector field $\e{_{(0)}}$ is by decomposing  $\nablab_\mu\e{_{(0)}}$ into its irreducible parts with respect to the rotation group \cite{Ehlers1993,Ellis2009}. This decomposition can be written as (see Sec.~II. D. of Ref~\cite{PhysRevD.108.044043})
\begin{align}
\rsconnection{_a_b_{(0)}}=\frac{1}{c}\left(\omega_{ab}+\theta_{ab}\right)+\frac{1}{c^2}a_a\delta^0_b,
\label{28112022e}
\\
\theta_{ab}=\sigma_{ab}+\frac{1}{3}\theta h_{ab},
\label{28112022f}
\end{align}
where\footnote{We have omitted the ``parentheses'' in the deltas for convenience, i.e., $\delta_a^{(0)}=\delta_a^{0}$.} $h_{ab}=\delta^i_a\delta^i_b=\delta^1_a\delta^1_b+\delta^2_a\delta^2_b+\delta^3_a\delta^3_b$,  $a_a$ is the acceleration, and $\sigma_{ab}$ is the shear tensor; the expansion and vorticity tensors are given by
\begin{align}
\theta_{(i)(j)}=\frac{c}{2}\left(\rsconnection{_{(i)(j)(0)}}+\rsconnection{_{(j)(i)(0)}}\right)
\label{28112022g}
\end{align}
(Note that $\theta=c\rsconnection{_{(i)(i)(0)}}$.) and  
\begin{align}
\omega_{(i)(j)}=\frac{c}{2}\left(\rsconnection{_{(i)(j)(0)}}-\rsconnection{_{(j)(i)(0)}}\right),
\label{28112022h}
\end{align}   
respectively.

From Eq.~\eqref{02062024b}, we see that $\rsconnection{_a_b_{(0)}}=\sqrt{2}m\hat{s}_{[a}\hat{\phi}_{b]}$. Therefore, the expansion tensor vanishes and the vorticity is simplify $\omega_{(i)(j)}=\sqrt{2}mc\hat{s}_{[(i)}\hat{\phi}_{(j)]}$, which yields $\omega_{(1)(2)}=mc/\sqrt{2}$. We may conclude from this result that the particles of the frame are rotating together with spacetime.

\subsubsection{Angular momentum}
To calculate $L^{ab}$, we need the superpotential, which can be calculated from Eq.~\eqref{10092020a}. Using Eqs.~\eqref{10092020a} and \eqref{02062024b}, we find
\begin{align}
\potential{_a_b_c}=\frac{m}{H}( \hat{t}_a\hat{t}_{[b}\hat{\phi}_{c]}+\hat{z}_a\hat{\phi}_{[b}\hat{z}_{c]} )+\frac{m}{\sqrt{2}}(\hat{s}_a\hat{t}_{[b}\hat{\phi}_{c]}-\hat{\phi}_a\hat{t}_{[b}\hat{s}_{c]}).
\label{17012025a}
\end{align}
Substitution into Eq.~\eqref{05062024c} yields
\begin{align}
M^{ab\mu}=-4km[\hat{t}^\mu(\hat{t}^{[a}\hat{\phi}^{b]}+\sqrt{2}H\hat{s}^{[a}\hat{\phi}^{b]})
\nonumber\\
+\frac{H}{\sqrt{2}}(-\hat{s}^\mu\hat{t}^{[a}\hat{\phi}^{b]}+\hat{\phi}^\mu\hat{t}^{[a}\hat{s}^{b]})
\nonumber\\
+\hat{z}^\mu\hat{\phi}^{[a}\hat{z}^{b]}].
\label{03062024a}
\end{align}
Using the expressions for the derivatives of the components of $\{\hat{t},\hat{s},\hat{\phi},\hat{z}\}$ given in Sec.~\ref{04032025a}, it is straightforward to show that $\partial_\mu M^{ab\mu}=0$, as expected.

Taking $\mu=0$ in \eqref{03062024a} and using Eq.~\eqref{02062024e}, we obtain
\begin{align}
M^{ab}=-4km H(\hat{t}^{[a}\hat{\phi}^{b]}+\sqrt{2}\hat{s}^{[a}\hat{\phi}^{b]}).
\label{03062024b}
\end{align}
Another way of obtaining $M^{ab}$ is to use Eq.~\eqref{19052024a}, which has the advantage of giving the result already in the form of a derivative. From Eqs.~\eqref{04032025b}, \eqref{02062024d}, \eqref{02062024e}, and the determinant of the tetrad field
\begin{align}
e=H(x),
\label{04032025c}
\end{align}
we obtain
\begin{align}
M^{ab}=-4k\partial_x[H\hat{t}^{[a}\hat{\phi}^{b]}-\sqrt{2}(1-H)\hat{s}^{[a}\hat{\phi}^{b]}].
\end{align}

If we integrate $M^{ab}$ in the region $x_1\leq x\leq x_2$, $y_1\leq y\leq y_2$, and $z_1\leq z\leq z_2$, and use Eq.~\eqref{30122019c}, we obtain
\begin{align}
L^{ab}=4k[H(x_2)-H(x_1)]A(\hat{t}^{[a}\hat{\phi}^{b]}+\sqrt{2}\hat{s}^{[a}\hat{\phi}^{b]}),
\label{09012025d}
\end{align}
where $A= \Delta y\Delta z$ is the area of $\{t=\textrm{constant},\ x=0,\ y_1\leq y\leq y_2,\ z_1\leq z\leq z_2\}$, which is obtained as follows: the area of the two-surface defined by constant  $t$ and $x$ is $A=\int\int dy dz \sqrt{\det g_{i'j'}}$, where $i',j'=2,3$; therefore  
\begin{align}
A(x)=\Delta y\Delta z \sqrt{2-[H(x)-2]^2},
\label{09012025c}
\end{align}
which yields $A(0)=\Delta y\Delta z$ [see Eq.~\eqref{09012025b}]. If we define the angular momentum per se as\footnote{The $c$ is to ensure that $L_d$ has the dimension of angular momentum.} $L_d=(1/2)L^{ab}\varepsilon_{(0)abd}/c$, we will find that it points along $z$, as expected.

The values of $x_1$ and $x_2$ cannot be arbitrary in the expression for $A(x)$, they have to satisfy $\ln(2-\sqrt{2})\leq mx\leq \ln(2+\sqrt{2})$. Let us evaluate $L_d$  using  $mx_1=\ln(2-\sqrt{2})$ and $mx_2=\ln(2+\sqrt{2})$. From Eq.~\eqref{09012025b}, we find $H(x_2)-H(x_1)=2\sqrt{2}$. Substituting this into Eq.~\eqref{09012025d} and calculating $L_d$, we arrive at $L_a=(8kA/c)\hat{z}_a$, which seems to be a consistent prediction for the angular momentum.

\subsubsection{Problems with $P^a$ in this TF}\label{26032025b}
Here we show that the tetrad field given by Eqs.~\eqref{02062024c} and  \eqref{02062024e} yields an inconsistent result for $P^a$. The possible cause is discussed at the end of this section.

Let us calculate $P^a$ in the same region as before. Since $e$ and $\potential{^a^0^i}$ depend only on $x$, the only contribution to $P^a$ in Eq.~\eqref{29102023c} will come from $\potential{^a^0^1}$. In turn, from Eq.~\eqref{17012025a}, we find that $\potential{^a^0^1}=-(m/2H)\hat{t}^a-(\sqrt{2}m/4)\hat{s}^a$. Using this equation and $e=H(x)$ in Eq.~\eqref{29102023c}, we obtain  $P^{(0)}=0$ and $P^{(i)}\neq 0$.  This is clear inconsistent.

What is the cause of this inconsistency? Why a frame that seems to be free from artificial properties and free from nongravitational interactions yield such a strange result? A possible answer is this. The frame we have used is not in the time gauge, as we can see from Eqs.~\eqref{02062024c} and \eqref{02062024e}: $\hat{t}_\mu=\e{_a_\mu}\hat{t}^a=\e{_{(0)}_\mu}$ leads to $\e{^{(0)}_\mu}=\delta^0_\mu+\sqrt{2}(H-1)\delta^2_\mu$. When Schwinger established the time gauge (see Ref.~\cite{PhysRev.130.1253} for details) the goal was to lock the time axes of the local coordinate systems to the time axis of the general coordinate system. This creates a relation between the hypersurface of constant $x^0$ and the tetrad field in such a way that the triad $\e{_{(k)}}$ lies inside the hypersurface and $\teta{^{(0)}}=\e{^{(0)}_\mu}dx^\mu$ is proportional to $dx^0$. This relation may perhaps be necessary\footnote{It is certainly not sufficient. See, for example, the discussion in Sec.~3.1 of Ref.~\cite{doi:10.1142/S0217732322502224}.} for consistency. 

The following is a list of possible restrictions on the TF (it is not exhaustive) that has guided authors to obtain consistent results.
\begin{enumerate}
\item\label{17012025c} $\rsconnection{^a_b_c}$ must vanish as the physical parameters go to zero. (This is the same as demand that the frame becomes a global inertial frame in Minkowski.)
\item\label{17012025d} No artificial properties, i.e., any change of orientation or in the clocks synchronization must be a consequence of either changes in the state of motion of the observers or due to the presence of a gravitational field. [This prevents local rotations in the surfaces of constant $x^0$, which is in agreement with the restriction imposed by the condition $\delta\e{^a_\mu}|_{\partial\mathcal{M}}=0$ on the boundary surface; this is important for the action principle (see, e.g., Refs.~\cite{Fiorini_2023}).]
\item\label{17012025e} $\e{_a^\mu}$ must satisfy the time gauge, i.e. $\e{_{(k)}^0}=0$ (equivalently $\e{^{(0)}_k}=0$), in a special type of coordinate system:
	\begin{enumerate}
	\item\label{17012025f} A Cartesian-like coordinate system  such that $\e{_{(i)j}}=\e{_{(j)i}}$ \cite{PhysRevD.65.124001}. (One may, perhaps, relax this condition by requiring $\e{_{(i)j}}$ to satisfy it only in the limit as the physical parameters go to zero.)
	\item\label{17012025g} And/or demand the coordinates to have finite values when the physical parameters go to zero. [For example, if $r$ is the Schwarzschild radial coordinate and the coordinate where the time gauge is satisfied is $R=Mr$, with $M$ being the black hole mass, then the frame may not be a good one even if it satisfies condition \ref{17012025c}. (Notice that $R\to 0$ as $M\to 0$ because $r$ has to be finite.) This is so because the frame is somewhat adapted to this coordinate, which means that the lack of a well-defined limit for the coordinates may render the frame problematic in some situations.]
	\end{enumerate}
\item\label{17012025b} The acceleration tensor must vanish, $\rsconnection{^a_{(0)}_b}=0$. (This may not be necessary, but it seems to help the understanding of the gravitational energy because the TF is free from nongravitational interactions.)
\end{enumerate}
Some consistent results may be obtained without imposing all those conditions.

The difficulty with the above conditions is that finding a tetrad field that satisfies all of them is hard, at least for most spacetimes. For example, the tetrad of Sec.~\ref{19012025a} does not satisfy condition  \ref{17012025e} and, possibly, condition \ref{17012025d} (verifying  this condition is not trivial). Another example is the tetrad adapted to the Gaussian coordinate system for G{\"o}del's universe given by Ref.~\cite{Novello1993}, which we discuss next. 

\subsection{Frame adapted to Gaussian coordinates}
In this section we discuss the difficulties in finding a ``good'' tetrad field adapted to Gaussian coordinates.  

The metric (18) of Ref.~\cite{Novello1993} can be written in the form\footnote{We set $c=1$.}
\begin{align}
ds^2=-d\tau^2+(\mu^2-1)d\lambda^2-a^2gd\eta^2
-2ah\,d\lambda d\eta+dZ^2,
\label{19012025bb}\\
g(r)=-\sinh^2r(1-\sinh^2r),\ h(r)=\sqrt{2}\sinh^2r,
\label{19012025b}
\\
1-\sin M=2\frac{\mu^2+1}{\mu^2-1}\sinh^2r,\ M=\frac{2}{a}\sqrt{\mu^2+1}(\tau-\mu\lambda),
\label{19012025bc}
\end{align}
where $r=r(\tau,\lambda)$, $M=M(\tau,\lambda)$, and $\mu$ is a parameter that labels different coordinate systems; the value of $\mu$ establishes   the range of $r$, which is given by $0\leq r\leq r_c$ with $\sinh^2r_c=(\mu^2-1)/(\mu^2+1)$ (note that $\mu^2>1$). It is worth noting that we are using slightly different coordinates\footnote{The signature is also different from that of Ref.~\cite{Novello1993}.}: $\tau\equiv\tilde{t}$, $\lambda\equiv a\tilde{\xi}$, $\eta\equiv\tilde{\eta}$ and $Z\equiv a\tilde{z}$. [See Eq.~(18) of Ref.~\cite{Novello1993} for comparison.] 

Although this Gaussian coordinates allow us to find a frame that is freely falling, with synchronized clocks and satisfying the time gauge, it is hard to find one that, in addition, becomes a global inertial frame of reference when $a$ goes to infinity. (The parameter $a$ is related to the physical parameter $\Omega$, the vorticity, through $a=\sqrt{2}/\Omega$. So, $a\to\infty$ implies $\Omega\to 0$.) To see this,  consider the coordinate transformation \cite{Novello1993}
\begin{align}
\tau=\mu T+\frac{a}{2}\sqrt{\mu^2+1}\arcsin\Psi+\frac{\mu a}{\sqrt{2}}\arcsin\Delta
\label{19012025c}
\\
\lambda=T+\frac{\mu a}{2\sqrt{\mu^2+1}}\arcsin\Psi+\frac{a}{\sqrt{2}}\arcsin\Delta
\label{19012025d}
\\
\eta=\phi-\frac{\pi}{4}+\frac{1}{2}\arcsin\Delta
\label{19012025e}
\\
Z=Z
\end{align}
and
\begin{align}
\Psi=1-2\frac{\mu^2+1}{\mu^2-1}\sinh^2(R/a)
\label{19012025f}
\\
\Delta=\frac{3\mu^2+1}{\mu^2-1}\frac{\sinh^2(R/a)}{\sinh^2(R/a)+1}-\frac{1}{\sinh^2(R/a)+1},
\label{19012025g}
\end{align}
where $R\equiv ar$. The coordinates $T$, $R$, $\phi$ and $Z$ have well-defined values in the limit $a\to\infty$ (vanishing physical parameter). However,  they become cylindrical coordinates in the absence of gravity. [Notice that $R$ is a radial coordinate. This can be seen by taking the limit of the metric \eqref{19012025bb} as $a\to\infty$.] This means that the coordinates $(\tau,\lambda,\eta,Z)$ are related to curvilinear coordinates, not with Cartesian-like coordinates. In this case, we have to find a tetrad that does not follows this curvilinear pattern. Because of the intricate relation between  $(\tau,\lambda,\eta,Z)$ and $(T,R,\phi, Z)$,  finding such a tetrad is not an easy task. 

The most natural choices to any tetrad field adapted to $(\tau,\lambda,\eta,Z)$ are the ones in which $\teta{^{(0)}}=d\tau$. They all satisfy the time gauge in the coordinates $(\tau,\lambda,\eta,Z)$. However,  after taking the limit $a\to\infty$ with $(T,R,\phi,Z)$ finite and $\mu$ fixed, we obtain $\lim_{a\to\infty}\teta{^{(0)}}=\mu dT+(\mu^2-1)^{1/2}dR$ (Remember that $\mu^2>1$). So, even in Minkowski spacetime, any tetrad with $\teta{^{(0)}}=d\tau$ will not satisfy the time gauge in either the coordinates $(T,R,\phi, Z)$ or $(T,X,Y, Z)$, where $X=R\cos\phi$ and $Y=R\sin\phi$. Furthermore,   neither $\tau$ nor $\lambda$ is well defined in the limit $a\to\infty$. Therefore, these tetrads do not satisfy condition \ref{17012025e}.

\subsection{Accelerated frame}\label{20012025n}
In order to analyze the consistency of $P^a$ in a tetrad field that satisfies condition \ref{17012025e}, let us ignore condition \ref{17012025b} and use an accelerated frame.  

G{\"o}del's universe can also be written as (see, e.g., footnote 2 in Ref.~\cite{Novello1993})
\begin{align}
ds^2=-dT^2+dR^2+dZ^2-2ah\, dT d\phi-a^2gd\phi^2,
\label{19012025h}
\end{align}
where $g$ and $h$ are given by Eq.~\eqref{19012025b}, with $r=R/a$. We choose our tetrad field to be
\begin{align}
\hat{t}^a=\delta^a_{(0)},\ \hat{x}^a=\delta^a_{(1)},\ \hat{y}^a=\delta^a_{(2)},\ \hat{z}^a=\delta^a_{(3)}
\label{20012025a}
\\
\hat{s}^a=\cos\phi\hat{x}^a+\sin\phi\hat{y}^a,\ \hat{\phi}^a=-\sin\phi\hat{x}^a+\cos\phi\hat{y}^a,
\label{20012025b}
\\ 
\hat{t}_\mu=-\sqrt{1+f^2}\delta^0_\mu,\ \hat{s}_\mu=\delta_\mu^1,\ \hat{\phi}_\mu=-f\delta_\mu^0+\alpha\delta_\mu^2,
\label{20012025c}\\
 \hat{z}_\mu=\delta_\mu^3,\ f\equiv \beta/\alpha,\ \alpha\equiv a\sqrt{-g},\ \beta\equiv ah, 
\label{20012025d}
\end{align}
where $\e{^a_\mu}$ is given in the same form as that of Eq.~\eqref{02062024c}. (Note, however, that $\hat{s}$ and $\hat{\phi}$ here are different; the coordinate $\phi$ is not the same either.) The scalar product of the unit vectors $\{\hat{t},\hat{s},\hat{\phi},\hat{z}\}$ are
\begin{align}
\hat{t}\cdot\hat{t}=-1,\ \hat{s}\cdot\hat{s}=\hat{\phi}\cdot\hat{\phi}=\hat{z}\cdot\hat{z}=1,
\label{13082024dd}\\
\hat{t}\cdot\hat{s}=\hat{t}\cdot\hat{\phi}=\hat{t}\cdot\hat{z}=\hat{s}\cdot\hat{\phi}=\ldots=0.
\label{13082024ee}
\end{align}
The components given by Eqs.~\eqref{20012025a} and \eqref{20012025b} are general, they can be used in any spacetime and do not fix the tetrad. The specialization to G{\"o}del's universe was made when we chose  Eqs.~\eqref{20012025c} and \eqref{20012025d}.

The tetrad we have chosen is not defined over the whole range of $r$ because we have assumed $-g>0$, which is not necessary for the coordinates in Eq.~\eqref{19012025h}. It is clear though that this tetrad satisfies the time gauge, since $\e{^{(0)}_\mu}=-\hat{t}_\mu=\sqrt{1+f^2}\delta^0_\mu$. In turn, the limit of $\teta{^a}=\e{^a_\mu}dx^\mu$ as $a\to\infty$ yields $\teta{^a}=(dT,dX,dY,dZ)$, where $X=R\cos\phi$ and $Y=R\sin\phi$, in agreement with the less restrictive version of condition \ref{17012025f}. This tetrad also satisfies conditions \ref{17012025g} and \ref{17012025c}; we hope it satisfies \ref{17012025d} as well, or at least does not deviate too much. Unfortunately, it does not satisfy the condition of being free from nongravitational fields, as we will see later.

In order to calculate the teleparallel quantities, we need $\torsion{^a_\mu_\nu}$ first. As before, after obtaining these components, we will only need to perform scalar products and contractions. It is thus convenient to write everything in terms of $\{\hat{t},\hat{s},\hat{\phi},\hat{z}\}$ as before; the only difference is that, now, we use the following identity\footnote{Equation \eqref{20012025i} can be easily verified by using the scalar products in Eqs.~\eqref{13082024dd} and \eqref{13082024ee}. In turn, Eqs.~\eqref{20012025e}-\eqref{20012025h} can be obtained by substituting Eq.~\eqref{02062024c} in Eq.~\eqref{04102019p} and comparing the result with Eq.~\eqref{20012025i}.}
\begin{align}
\torsion{^a_\mu_\nu}=-\torsion{^{\hat{t}}_\mu_\nu}\hat{t}^a+\torsion{^{\hat{s}}_\mu_\nu}\hat{s}^a+\torsion{^{\hat{\phi}}_\mu_\nu}\hat{\phi}^a+\torsion{^{\hat{z}}_\mu_\nu}\hat{z}^a,
\label{20012025i}
\end{align}
where
\begin{align}
\torsion{^{\hat{t}}_\mu_\nu}=2\partial_{[\mu}\hat{t}_{\nu]},
\label{20012025e}
\\
\torsion{^{\hat{s}}_\mu_\nu}=2\partial_{[\mu}\hat{s}_{\nu]}-2\delta_{[\mu}^2\hat{\phi}_{\nu]},
\label{20012025f}
\\
\torsion{^{\hat{\phi}}_\mu_\nu}=2\partial_{[\mu}\hat{\phi}_{\nu]}+2\delta_{[\mu}^2\hat{s}_{\nu]},
\label{20012025g}
\\
\torsion{^{\hat{z}}_\mu_\nu}=2\partial_{[\mu}\hat{z}_{\nu]}.
\label{20012025h}
\end{align}
These identities are written in the cylindrical-like coordinate system $(T,R,\phi,Z)$. They are the analog of Eqs.~(28)-(31) of Ref.~\cite{Formiga2021Braz}, which are written in spherical coordinates\footnote{Notice that Eqs.~(28)-(31) of Ref.~\cite{Formiga2021Braz} will have the same form as that of Eqs.~\eqref{20012025e}-\eqref{20012025h} if we use the notation $\delta_\mu^\phi$, rather than $\delta_\mu^3$ for spherical coordinates and $\delta_\mu^2$ for cylindrical ones.}. They are useful because they simplify the calculations and can be used for any spacetime, as long as the coordinate system is of the form $x^\mu=(x^0,x^1,x^2=\phi,x^3)$. 
 
After applying the partial derivatives $\partial_\mu$ in the Eqs.~\eqref{20012025c}-\eqref{20012025d}, we use those equations again to eliminate the deltas that appear. (For example, $\partial_\mu f=(df/dR)\delta_\mu^1=f'\hat{s}_\mu$.)  This gives
\begin{align}
\torsion{^a_\mu_\nu}=&\ 2F_1\hat{t}_{[\mu}\hat{s}_{\nu]}\hat{t}^a+2F_2\hat{t}_{[\mu}\hat{\phi}_{\nu]}\hat{s}^a+
\nonumber\\
&+2\left(F_3\hat{t}_{[\mu}\hat{s}_{\nu]}+F_4\hat{s}_{[\mu}\hat{\phi}_{\nu]}\right)\hat{\phi}^a,
\label{20012025j}
\end{align} 
where
\begin{align}
F_1=\frac{ff'}{1+f^2},\ F_2=\frac{f}{\alpha\sqrt{1+f^2}},
\label{20012025k}\\
F_3=-\frac{[f'+(1-\alpha')f/\alpha]}{\sqrt{1+f^2}},\ F_4=\frac{-(1-\alpha')}{\alpha}.
\label{20012025l}
\end{align}
(The prime denotes differentiation with respect to $R$.) Substitution into Eq.~\eqref{10092020a} gives
\begin{align}
\rsconnection{_a_b_c}=&-\left[2F_1\hat{t}_b+(F_2+F_3)\hat{\phi}_b\right]\hat{t}_{[a}\hat{s}_{c]}
\nonumber\\
&+\left[(F_3-F_2)\hat{t}_b-2F_4\hat{\phi}_b\right]\hat{s}_{[a}\hat{\phi}_{c]}
\nonumber\\
&-(F_2+F_3)\hat{s}_b\hat{t}_{[a}\hat{\phi}_{c]},
\label{20012025m}
\end{align}
from which we find $a_b=\phi_{(0)b}=F_1\hat{s}_b$ and $\omega_{a}=(1/2)(F_3-F_2)\hat{z}_a$, where $\omega_a=(1/2)\varepsilon_{(0)abc}\phi^{bc}$; $\varepsilon_{dabc}$ is the Levi-Civita tensor. Thus, the frame is accelerated along $\hat{s}$ and rotates about $\hat{z}$.

In calculating the energy we need $\potential{^a^0^i}$ and $e$. The latter is
\begin{align}
e=\sqrt{\alpha^2+\beta^2},
\label{21012025c}
\end{align}
while the former can be obtained from Eq.~\eqref{14082024h} and \eqref{20012025m}. Using $\rsconnection{^a_a_b}=(F_1+F_4)\hat{s}_b$ and $\eta_{ab}=-\hat{t}_a\hat{t}_b+\hat{s}_a\hat{s}_b+\hat{\phi}_a\hat{\phi}_b+\hat{z}_a\hat{z}_b$, we find
\begin{align}
\potential{^a^b^c}=&\ \left[-F_4\hat{t}^a+\frac{1}{2}(F_2+F_3)\hat{\phi}^a\right]\hat{t}^{[b}\hat{s}^{c]}
\nonumber\\
&+\left[\frac{1}{2}(F_2-F_3)\hat{t}^a-F_1\hat{\phi}^a\right]\hat{s}^{[b}\hat{\phi}^{c]}
\nonumber\\
&+\frac{1}{2}(F_2+F_3)\hat{s}^a\hat{t}^{[b}\hat{\phi}^{c]}-(F_1+F_4)\hat{z}^a\hat{s}^{[b}\hat{z}^{c]},
\label{21012025d}
\end{align}
which, with the help of Eq.~\eqref{20012025o}, leads to
\begin{align}
\potential{^a^0^i}=&\ \frac{1}{2}\left[-F_4\hat{t}^a+\frac{1}{2}(F_2+F_3)\hat{\phi}^a\right]\frac{\delta^i_1}{\sqrt{1+f^2}}
\nonumber\\
&+\frac{(F_2+F_3)\hat{s}^a\delta^i_2}{4\alpha\sqrt{1+f^2}}.
\label{21012025e}
\end{align}
Substituting Eqs.~\eqref{21012025c} and \eqref{21012025e} into Eq.~\eqref{29102023c} and integrating over the surface of a cylinder, we obtain
\begin{align}
P^a=4\pi k\Delta Z (1-\alpha')\hat{t}^a.
\label{21012025f}
\end{align}
Although this expression diverges when $\sinh^2(R/a)=1$, where the tetrad becomes problematic, it does not yield $E=0$ and $P^{(i)}\neq 0$. 

In some sense, an infinity energy seems more reasonable than $P^a=(0,\vec{P})$ because G{\"o}del's universe possesses a perfect fluid with a constant energy density over the whole universe, which can naturally lead to a divergence in the total energy. In the specific case considered here, this divergence is possibly caused by the acceleration of the frame. On the other hand, it seems that the result $P^a=(0,\vec{P})$ cannot be explained by any unphysical assumption about the dynamics of the frame or G{\"o}del's solution.

For completeness, we can evaluate the antisymmetric part of $\energy{^a^b}$. Using Eqs.~\eqref{20012025j} and \eqref{21012025d} in Eq.~\eqref{24122024a}, we find that
\begin{align}
2\energy{^{[a}^{b]}}=-2k(F_1+F_4)(F_2+F_3)\hat{t}^{[a}\hat{\phi}^{b]}.
\end{align}
Hence, the gravitational energy-momentum tensor is not symmetric.

\section{Concluding remarks}\label{26032025a}
We have obtained expressions \eqref{20122024a} and \eqref{23122024a}, and also their versions in a TF, Eqs.~\eqref{07012025b} and \eqref{29102023d}; this generalizes the angular momentum approach in the TEGR and, in some sense, complete the analogy with special relativity.

We have discussed the interpretation of $L^{ab}$ and conclude that it is unlikely that it represents an angular momentum associated to the gravitational field, and that it is also unlikely to represent an ordinary orbital angular momentum; it is more likely connected to a kind of intrinsic angular momentum of the matter fields.

We have specialized our analysis to G{\"o}del's universe and given examples with many different tetrads. The results suggest that the time gauge is essential for $P^a$. Although the vanishing of $L^{(i)(j)}$ in the time gauge is not necessarily inconsistent, because its physical meaning is still obscure, this opens the question of how one can define an angular momentum that can be compatible with $P^a$ and, at the same time, be nontrivial in spacetimes such as G{\"o}del's.

Perhaps, $L^{(i)(j)}$ vanishes in all spacetimes, including $pp$-wave spacetimes with circular polarization (see, e.g., Ref.~\cite{doi:10.1142/S0217732325501056}), Kerr, G{\"o}del etc. because it is not an orbital angular momentum and its intrinsic nature cannot be properly revealed in a classical theory. (See Refs.~\cite{PhysRevD.67.044016,PhysRevD.67.108501} for a discussion of the coupling with spin.) Of course, there is always the possibility that one cannot relate $L^{ab}$ to any physical quantity in a consistent way.

\appendix

\section{Derivatives of $\{\hat{t},\hat{s},\hat{\phi},\hat{z}\}$ for Sec.~\ref{19012025a}}\label{04032025a}
Here we exhibit some useful expressions for the derivatives of the components of the basis $\{\hat{t},\hat{s},\hat{\phi},\hat{z}\}$.

From Eqs.~\eqref{02062024d} and \eqref{02062024e}, one finds that
\begin{align}
\partial_\mu\hat{t}^a=\partial_\mu\hat{z}^a=0,\ \partial_\mu\hat{s}^a=\frac{m}{\sqrt{2}}\hat{\phi}^a[\hat{t}_\mu+\sqrt{2}(1-\frac{1}{H})\hat{s}_\mu],
\nonumber\\
\partial_\mu\hat{\phi}^a=-\frac{m}{\sqrt{2}}\hat{s}^a[\hat{t}_\mu+\sqrt{2}(1-\frac{1}{H})\hat{s}_\mu].
\label{03032025a}
\end{align}
and 
\begin{align}
\partial_\mu\hat{t}_\nu=\sqrt{2}m\hat{\phi}_\mu\hat{s}_\nu,\  \partial_\mu\hat{s}_\nu=-m\hat{\phi}_\mu\hat{s}_\nu,
\nonumber\\
\partial_\mu\hat{\phi}_\nu=\partial_\mu\hat{z}_\nu=0.
\label{03032025b}
\end{align}

If necessary, one can also use
\begin{align}
\partial_\mu\hat{t}^\nu=\partial_\mu\hat{\phi}^\nu=\partial_\mu\hat{z}^\nu=0,
\\
\partial_\mu\hat{s}^\nu=m\hat{\phi}_\mu(\sqrt{2}\hat{t}^\nu+\hat{s}^\nu).
\end{align}

\section{Calculations of Sec.~\ref{20012025n}}
The inverse metric of Eq.~\eqref{19012025h} is
\begin{align}
g^{\mu\nu}=&\ \frac{-\delta^\mu_0\delta^\nu_0}{1+f^2}+\delta^\mu_1\delta^\nu_1+\frac{\delta^\mu_2\delta^\nu_2}{\alpha^2(1+f^2)}+\delta^\mu_3\delta^\nu_3
\nonumber\\
&-\frac{f}{\alpha(1+f^2)}(\delta^\mu_0\delta^\nu_2+\delta^\mu_2\delta^\nu_0),
\end{align}
where $f$ is given by Eq.~\eqref{20012025d}. Now we can raise the indices of Eqs.~\eqref{20012025c} and \eqref{20012025d}:
\begin{align}
\hat{t}^\mu=\frac{\delta^\mu_0+(f/\alpha)\delta^\mu_2}{\sqrt{1+f^2}},\ \hat{s}^\mu=\delta^\mu_1,\ \hat{\phi}^\mu=\frac{\delta^\mu_2}{\alpha},\ \hat{z}^\mu=\delta^\mu_3.
\label{20012025o}
\end{align}
In order to write everything in terms of the unit vectors, it is useful to invert Eqs.~\eqref{20012025c}, \eqref{20012025d} and \eqref{20012025o}:
\begin{align}
\delta_\mu^0=\frac{-\hat{t}_\mu}{\sqrt{1+f^2}},\ \delta_\mu^1=\hat{s}_\mu,\ \delta_\mu^2=\frac{-f\hat{t}_\mu}{\alpha\sqrt{1+f^2}}+\frac{\hat{\phi}_\mu}{\alpha},\ \delta_\mu^3=\hat{z}_\mu,
\label{21012025a}\\
\delta^\mu_0=(1+f^2)^{1/2}\hat{t}^\mu-f\hat{\phi}^\mu,\ \delta^\mu_1=\hat{s}^\mu,\ \delta^\mu_2=\alpha\hat{\phi}^\mu,\ \delta^\mu_3=\hat{z}^\mu.
\label{21012025b}
\end{align}

\section*{Acknowledgments}
Costa, R. D. acknowledges CAPES for financial support.


%

\end{document}